# Effect of Cu intercalation and pressure on excitonic interaction in 1*T*-TiSe$_2$

S. Kitou[1,2], A. Nakano[1,†], S. Kobayashi[1,‡], K. Sugawara[1,§], N. Katayama[1], N. Maejima[3,#],
A. Machida[3], T. Watanuki[3], K. Ichimura[4,5], S. Tanda[4,5], T. Nakamura[2], and H. Sawa[1]

[1]*Department of Applied Physics, Nagoya University, Nagoya 464-8603, Japan*
[2]*Institute for Molecular Science, Myodaiji, Okazaki 444-8585, Japan*
[3]*Synchrotron Radiation Research Center, National Institutes for Quantum and Radiological Science and Technology, Sayo, Hyogo 679-5148, Japan*
[4]*Department of Applied Physics, Hokkaido University, Sapporo 060-8628, Japan*
[5]*Center of Education and Research for Topological Science and Technology, Hokkaido University, Sapporo 060-8628, Japan*

1*T*-TiSe$_2$ has a semimetallic band structure at room temperature and undergoes phase transition to a triple-*q* charge density wave (CDW) state with a commensurate superlattice structure ($2a \times 2a \times 2c$) below $T_c \approx 200$ K at ambient pressure. This phase transition is caused by cooperative phenomena involving electron–phonon and electron–hole (excitonic) interactions, and cannot be described by a standard CDW framework. By Cu intercalation or the application of pressure, this phase transition temperature is suppressed and superconductivity (SC) appears. However, it is not clear what kind of order parameters are affected by these two procedures. We investigated the crystal structure of Cu$_x$TiSe$_2$ and pressurized 1*T*-TiSe$_2$ around the SC state by synchrotron x-ray diffraction on single crystals. In the high-temperature phase, the variation of structural parameters for the case of Cu intercalation and application of pressure are considerably different. Moreover, the relationship between the critical points of the CDW phase transition and the SC dome are also different for the two cases. The excitonic interaction appears to play an important role in the *P*−*T* phase diagram of 1*T*-TiSe$_2$, but not in the *x*−*T* phase diagram.

## I. INTRODUCTION

In strongly correlated electronic systems, exotic electronic states are often realized by changing the balance between interactions through external pressure and/or carrier doping. For example, superconductivity (SC) in cuprates appears by suppressing the antiferromagnetic interaction through carrier doping [1]. Hence, in these materials, it is very important to understand which of the fundamental parameters change by applying external pressure and carrier doping. The electronic state of transition metal dichalcogenides (TMDs), which changes on carrier doping or application of external pressure, is interesting from this point of view.

TMDs often show various charge density wave (CDW) states, with SC often appearing in its vicinity in materials such as 1*T*-TaS$_2$ [2,3] and 2*H*-NbSe$_2$ [4,5]. In this series, 1*T*-TiSe$_2$ has been one of the most vigorously researched systems because of its exotic electronic states [6-10] such as chiral CDW [11], and exciton condensation [12]. Although 1*T*-TiSe$_2$ has both electron–phonon and electron–hole interactions, several aspects of its electronic ground state remain to be understood. In this compound, the formal valence states of Ti$^{4+}$ ($3d^0$) and Se$^{2-}$ ($4p^6$) correspond to closed shells. According to reports, a semimetallic band structure is formed by Se $4p$ and Ti $3d$ orbitals [13]. Near the Fermi energy, a hole pocket of the Se $4p$ and electron pockets of the Ti $3d$ exist at the *Γ* point and the *L* points, respectively. The electrical resistivity shows metallic characteristics at room temperature [14].

By decreasing temperature, a hump appears in electric resistivity at $T_c \approx 200$ K [14]. Below 200 K (*β* phase), a triple-*q* CDW state with a commensurate superlattice structure ($2a \times 2a \times 2c$) is formed by folding of the band near *L* points to *Γ* point [15-17]. However, because the Fermi surface in 1*T*-TiSe$_2$ is three-dimensional and the size of two Fermi surfaces contributing to the nesting between electrons and holes is different, the triple-*q* nesting condition from the hole-pocket to the electron-pockets ($q_{nest} = \boldsymbol{a}^*/2 + \boldsymbol{c}^*/2, -\boldsymbol{b}^*/2 + \boldsymbol{c}^*/2, -\boldsymbol{a}^*/2 + \boldsymbol{b}^*/2 + \boldsymbol{c}^*/2$) is not good. Therefore, the origin of this phase transition cannot be explained within a simple CDW framework. In this phase transition mechanism, it is argued that not only the electron–phonon coupling (EPC) [17-21] but also the electron–hole (excitonic) interaction [12,22-25] plays an important role. However, because the electronic system and the lattice system are strongly coupled in this compound, it is difficult to accurately estimate the contribution of the excitonic interaction exclusively. Ta$_2$NiSe$_5$, which is a direct gap semiconductor that has similarities to 1*T*-TiSe$_2$ [26,27], is considered another candidate as an excitonic insulator [28-33]. Comparing the electronic states of both compounds is important for understanding the excitonic insulator system.

Recently, SC was achieved in 1*T*-TiSe$_2$ by intercalation [34-37], external pressure [38], and electric field [9]. For example, SC was reported for $0.04 \leq x \leq 0.10$ ($T_{SC}^{MAX} = 4.15$ K at $x = 0.08$) in Cu intercalated Cu$_x$TiSe$_2$ [34], and in the pressure range of 2–4 GPa ($T_{SC}^{MAX}$ = 1.8 K at $P$ = 3 GPa) in pristine 1$T$-TiSe$_2$ [38]. These two SC states are summarized using similar pressure−temperature ($P$−$T$) and Cu content $x$−temperature ($x$−$T$) phase diagrams. It is prime interest to understand how the change of the electronic and/or the lattice system influences the triple-$q$ CDW state and the SC state in 1$T$-TiSe$_2$. Indeed, multilateral experiments and calculations are performed in Cu$_x$-TiSe$_2$ [39-49] and pressurized 1$T$-TiSe$_2$ [49-52].

The crystal structure of pristine 1$T$-TiSe$_2$ has been studied using neutron diffraction by DiSalvo *et al.* in 1976 [14], but detailed information about the crystal structure around $T_{SC}$ is hardly reported because of the experimental and analytical difficulties described later. The theoretical investigations of the SC state in 1$T$-TiSe$_2$ [47-49,52] were calculated by using the crystal structure of 1$T$-TiSe$_2$ or Cu$_x$TiSe$_2$ in the high-temperature (HT). Cu intercalation changes the chemical potential and pressurization changes the phonon modes by changing lattice structures. It is strange that the two phase diagrams of Cu intercalated [34,39] and pressurized [38,50] 1$T$-TiSe$_2$ resemble each other. Therefore, to understand the anomalous electronic state of 1$T$-TiSe$_2$, it is important to understand the structural changes associated with Cu intercalation and pressurization.

In this study, we conducted the synchrotron radiation x-ray diffraction (XRD) experiments and crystal structure analysis of 1$T$-TiSe$_2$ under ambient- and high-pressure conditions as well as Cu intercalated 1$T$-TiSe$_2$ under ambient-pressure. In Cu$_x$TiSe$_2$, detailed structural parameters and a complete phase diagram with the information of phase transitions and the SC dome are reported. Furthermore, for 1$T$-TiSe$_2$ under high-pressure, structural parameters were determined with a high degree of precision by using multiple single crystals and by performing structure analysis including superlattice reflections.

## II. EXPERIMENT

Single crystal samples of 1$T$-TiSe$_2$ used under ambient- and high-pressure experiments were synthesized using the procedure outlined in Ref. [14], and five kinds of single crystal samples of Cu$_x$TiSe$_2$ ($x$ = 0.05–0.13) were prepared with according to Ref. [53]. The content of Cu in Cu$_x$TiSe$_2$ was determined by crystal structure analysis using the synchrotron XRD at room temperature, in which weak diffuse scattering corresponding to disordered Cu position was ignored. The XRD measurements at ambient-pressure were carried out at beamline BL02B1 at the synchrotron facility SPring-8, Japan [54]. A helium gas blow was employed to cool the sample to 25 K at BL02B1. The XRD measurements under high-pressure were performed using a diamond anvil cell (DAC) apparatus at beamline BL22XU at SPring-8 [55], using a wavelength $\lambda$ = 0.4133 Å. The single crystals of 1$T$-TiSe$_2$ were loaded into a hole (220 μm diameter) in a stainless steel gasket. The incident x-ray beam was shaped into a 40 × 40 μm$^2$ square and impinged into the samples. Helium was used as the pressure medium and helium-gas-membrane system was used for pressurization. The pressure was calibrated by measuring the fluorescence of small rubies placed beside the 1$T$-TiSe$_2$ crystal in the sample chamber [56]. A refrigerator was employed to decrease the temperature to 5 K at BL22XU. A two-dimensional imaging-plate (IP) was used as the diffractometer's detector at the two beamlines to perform structure analysis including superlattice reflections. For the crystal structure analysis, we used original software for extracting the peak intensity [57]. Because the diffraction from the single crystals of diamond cannot be ignored in a single crystal XRD experiment using DAC, these are also taken into consideration for the analysis. Peak-intensity averaging and structure analysis were performed using SORTAV [58] and Jana2006 [59], respectively.

The electric resistivity measurements were performed in a Quantum Design physical properties measurement system (PPMS) instrument using the standard four-probe technique. Additional resistivity measurements down to 100 mK were carried out by using an adiabatic demagnetization refrigerator cell combined with the PPMS. DC magnetization measurements were conducted using a superconducting quantum interference device magnetometer in a Quantum Design magnetic properties measurement system instrument.

## III. RESULT
### A. 1$T$-TiSe$_2$ at ambient-pressure

Figure 1(a) shows the crystal structure of 1$T$-TiSe$_2$ in the HT phase ($a × a × c$; $P\bar{3}m1$). Ti and Se each form a triangular lattice. Below 200 K, reflections appeared at ($h$/2, $k$/2, $l$/2) (shown in Fig. 2), and a superlattice structure ($2a × 2a × 2c$; $P\bar{3}c1$) is realized [Fig. 1(b)] in the $\beta$ phase of 1$T$-TiSe$_2$. The structural parameters obtained from our XRD experiment at 30 K (Table I) are qualitatively consistent with the results reported by DiSalvo *et al.* [14]. As a result of the atomic displacements accompanying the phase transition, hybridization of Se 4$p$ and Ti 3$d$ orbitals creates the triple-$q$ structure [15-17]. These displacements also correspond to softening of the transverse optical (TO) phonon mode observed by the inelastic x-ray scattering (IXS) [20]. Each of the Ti and Se atoms has one symmetry site in the HT phase, whereas they are divided into two symmetry sites each in the $\beta$ phase. The ratio of the number of Ti and Se sites is 1:3 in the $\beta$ phase. As a result, TiSe$_6$ octahedra with symmetry $D_{3d}$ are divided into two sets of octahedra with symmetries $D_3$ and $C_2$ during the structural phase transition [the dotted circle and the solid circle shown in Fig. 1(b), respectively]. The volumes of the two TiSe$_6$ octahedra are nearly equal ($D_3$: 22.1402 Å$^3$, $C_2$: 22.1452 Å$^3$) at 30 K.

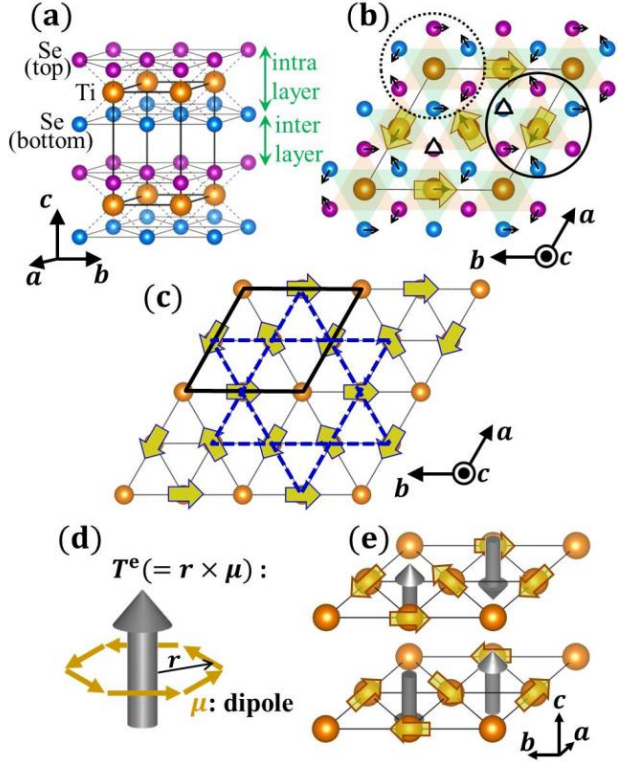

FIG. 1(a) Crystal structure of 1$T$-TiSe$_2$ ($a \times a \times c$). (b) Crystal structure ($2a \times 2a \times 2c$) in the $\beta$ phase. The black arrows on atoms show the atomic displacement accompanying the structural phase transition. The yellow arrows show the electric dipole moments. There is a three-fold axis along the $c$ axis at sites indicated by white triangles. The dotted circle and the solid circle show the octahedra with symmetries $D_3$ and $C_2$, respectively. (c) The electric dipole moments form a kagome lattice (blue dotted line). The black line shows a unit cell of size $2a \times 2a$. (d) Formation of an electric toroidal moment $T^e$. (e) Arrangement of the electric dipole moments $\mu$ and the electric toroidal moments $T^e$ in 1$T$-TiSe$_2$ in the $\beta$ phase.

In TiSe$_6$ octahedra with symmetry $C_2$, Ti and a part of Se are displaced in opposite directions. Therefore, the electric dipole moments exist locally despite the large electrical conductivity in the $\beta$ phase [14], whereas no dipole moments in TiSe$_6$ octahedra with symmetry $D_3$. These electric dipole moments have been calculated [60], and they correspond to the triple-$q$ structure [15-17], which was formed by the freezing of the TO phonon mode [20]. The intensity of the superlattice reflections is scaled to the size of the dipole moment $|\mu|$ ($\mu = q d$). From the difference between the centroids of Ti and Se atoms, the value of $|d|$ is 0.066 Å. In the case of BaTiO$_3$ (Ti$^{4+}$ 3$d^0$),
which is a typical ferroelectric material, the value of $|d|$ is 0.134 Å in the ferroelectric phase ($P4mm$) [61]. Hence, the $|\mu|$ in 1$T$-TiSe$_2$, which is about half value in BaTiO$_3$, is significant.

In this system, one-fourth of TiSe$_6$ octahedra become nodes that do not develop dipoles [the dotted circle in Fig. 1(b)]. As a result, the dipoles form the kagome lattice in the basal plane, as shown in Fig. 1(c). On this kagome lattice, the electric dipoles form vortices around the lattice sites indicated by white triangles, a three-fold rotation axis exists [Fig. 1(b)]. Because there is an equal number of clockwise and anticlockwise vortices, antiferroelectric arrangement of electric dipoles are realized within the TiSe$_2$ plane. In the center of the vortex of the electric dipoles, an electric toroidal moment $T^e = r \times \mu$ can be defined as shown in Fig. 1(d) [62-65]. Because $r$ is a vector from the center of the vortex to the dipole moment $\mu$, $T^e$ is along the $c$ axis at the center of the vortex. An antiferrotoroidic state, where the neighboring electric toroidal moments are arranged in opposite directions both within and between layers, is realized [Fig. 1(e)]. Figure 2 shows the temperature dependence of lattice constants normalized to 300 K. The rate of change in the $c$ axis lattice constant is lower below $T_c$. This unusual behavior, consistent with the reported data [66], may be due to the antiferrotoroidic state in 1$T$-TiSe$_2$.

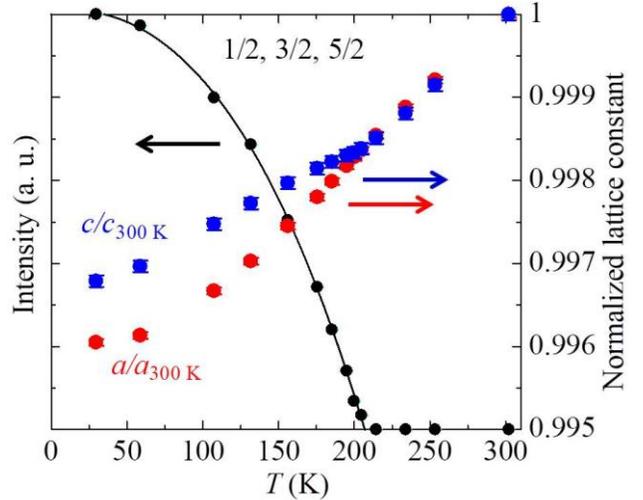

FIG. 2 Temperature dependence of the intensity of the (1/2, 3/2, 5/2) superlattice peak (black) and the lattice constants normalized to 300 K (red and blue).

## B. Cu$_x$TiSe$_2$ at ambient-pressure

The formation of electric dipoles is important in the triple-$q$ CDW state of pristine 1$T$-TiSe$_2$ because these dipoles correspond to the triple-$q$ structure, which is formed by the freezing of the TO phonon mode. Our interest here is to understand how these electric dipoles change by Cu intercalation or application of an external pressure.

For Cu intercalated samples, the lattice constants and $a/c$ at room temperature for Cu intercalated samples Cu$_x$TiSe$_2$ with different $x$ are shown in Fig. 3(a). By increasing $x$, the $c$ axis parameter increases appreciably, while the $a$ axis parameter also increases monotonically. There is a linear relationship between $a/c$ and $x$, as shown in the inset of Fig. 3(a). Compared to the results of previous work by Morosan et al. [34][34], nearly the same values of $a/c$ were observed for Cu content $x$ at room temperature. A linear equation

$$a/c = -2.237 \times 10^{-2} x + 0.589, \qquad (1)$$

can be obtained for approximating the linear relationship between $x$ and $a/c$ [black line in inset of Fig. 3(a)].

creases with increasing $x$ [Fig. 3(b)]. The volume of TiSe$_6$ octahedra increases with increasing $x$ [inset of Fig. 3(b)], indicating a decrease in the valence of Ti. This corresponds to electron doping from Cu ions to TiSe$_2$ layers, which has been also verified through the density-functional-theory calculations [48]. Assuming the valence of Se (2-) in $1T$-TiSe$_{2-\delta}$, since Se vacancy $\delta$ also corresponds to electron doping, $\delta$ dependence of the Ti-Se bond length similar to Cu$_x$TiSe$_2$ is predicted. However, the result of the structure analysis reported in Ref. [67] shows that the Ti-Se bond length decreases with increasing $\delta$, which is an inverse correlation with Cu intercalation system. The Se vacancy cannot be understood by the simple valence change of Ti.

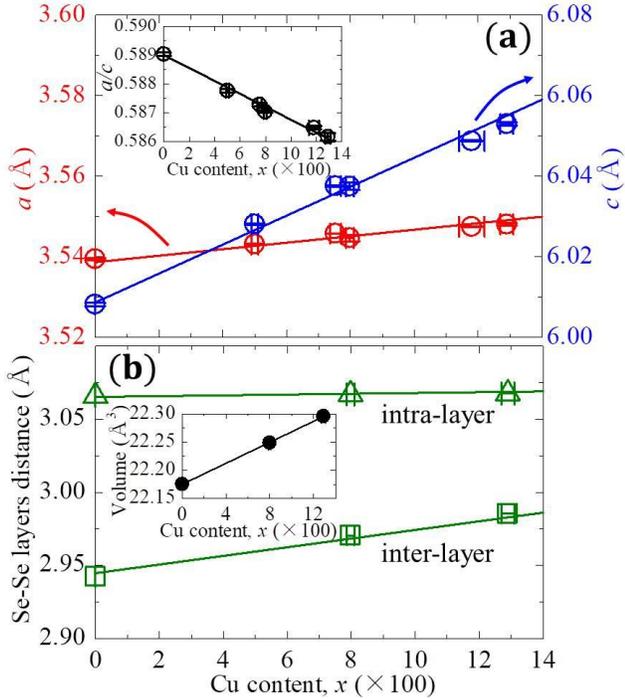

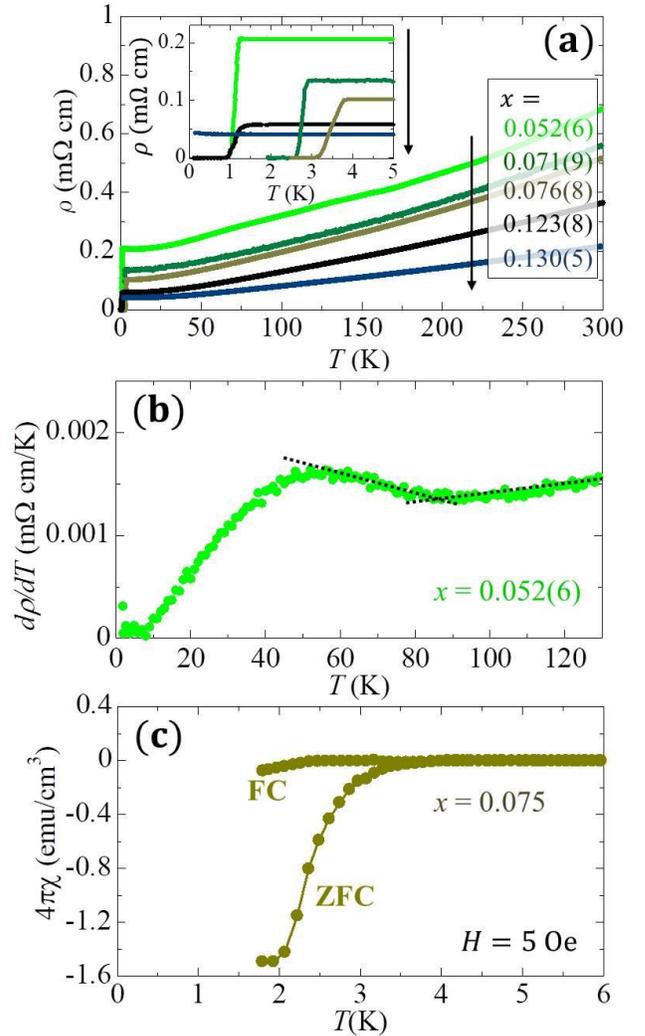

FIG. 3(a) Lattice parameters of Cu$_x$TiSe$_2$ with Cu content $x$ = 0, 0.0500(11), 0.0753(17), 0.0798(12), 0.118(4), and 0.129(2) at room temperature. Inset: $a/c$ as a function of $x$ at room temperature. (b) The distances Se-Se (intra-layer) and Se-Se (inter-layer) at room temperature for $x$ = 0, 0.0798(12), and 0.129(2) [refer to Fig. 1(a)]. Inset: Volume of TiSe$_6$ octahedra as a function of $x$ at room temperature.

Table I shows the resulst of the structure analysis for $x$ = 0, 0.0798(12), and 0.129(2) at room temperature. The Cu ion occupies the $1b$ site (Wyckoff letter) of $P\bar{3}m1$ space-group, which is the site between TiSe$_2$ layers. Since it has been reported that the physical properties are affected by the Se vacancy in $1T$-TiSe$_2$ [67], we also investigated this fact. We confirmed that there was no Se vacancy in all Cu$_x$TiSe$_2$ samples including non-doped $1T$-TiSe$_2$. The inter-layer Se-Se distance [Fig. 1(a)] in-

FIG. 4(a) Temperature dependence of the electric resistivity for Cu$_x$TiSe$_2$. Inset: Resistivity below 5 K. (b) $d\rho/dT$ curve for $x$ = 0.052(6), $T_c \approx$ 80 K. (c) Magnetic susceptibility of Cu$_{0.075}$TiSe$_2$.

Figure 4(a) shows the temperature dependence of the electric resistivity for Cu$_x$TiSe$_2$. The value $x$ of single crystal samples used for electric resistivity measurements was determined from XRD experiments by using Eq. (1). SC was confirmed in samples with $0.052(6) \leq x \leq 0.123(8)$, with $T_{SC}^{MAX} = 3.79$ K occurring at $x = 0.076(8)$, consistent with reported composition [34]. In $x = 0.130(5)$, the SC was absent down to 139 mK. Figure 4(c) shows the magnetic susceptibility of $Cu_{0.075}TiSe_2$, $T_{SC} = 3.8$ K, the SC volume fraction is large enough to constitute bulk SC.

The phase transition temperature was determined as around 80 K for $x = 0.052(6)$ from the $d\rho/dT$ curve [Fig. 4(b)]. Furthermore, the same superlattice pattern as pristine $1T$-$TiSe_2$ was confirmed in $Cu_{0.05}TiSe_2$ below 80 K from XRD data [Fig. 5(a) and Fig. 6(a)]. We confirmed that some superlattice peaks with the strong intensity show a broadening along $\bm{q}_{nest} = \bm{a}^*/2 \pm \bm{c}^*/2$ vectors in $Cu_{0.05}TiSe_2$ at 30 K [Fig. 5(a)]. For example, a superlattice reflection $(-5/2, 3, -1/2)$ extends in two directions of $-\bm{a}^*/2 + \bm{c}^*/2$ and $-\bm{a}^*/2 - \bm{c}^*/2$. These directions correspond to the $\bm{q}_{nest}$ vectors in two directions passing through an $L$ point [Fig. 5(d)]. The correlation length was calculated to be about 10 unit cells $\times \sqrt{7^2 + 12^2} \approx 140$ Å from the peak broadening in the $\bm{q}_{nest}$ direction of the superlattice reflections. This is the same order of magnitude as the domain size reported from earlier XRD [39] and scanning tunneling microscopy measurements [41].

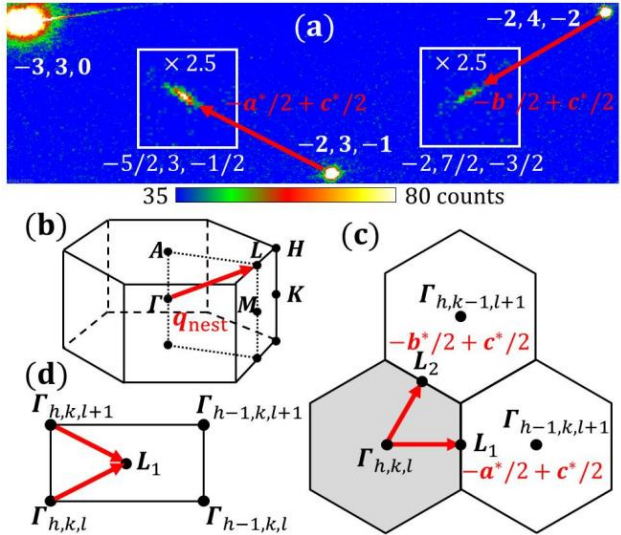

FIG. 5 (a) X-ray diffraction data of $Cu_{0.05}TiSe_2$ at 30 K. The white square parts are shown as 2.5 times as large. (b)–(d) Brillouin zone layout of $1T$-$TiSe_2$ at high-temperature phase ($a \times a \times c$) where $\Gamma$ (0, 0, 0), $A$ (0, 0, 1/2), $M$ (1/2, 0, 0), $L$ (1/2, 0, 1/2), $K$ (1/3, 1/3, 0), and $H$ (1/3, 1/3, 1/2). The red arrows show the $\bm{q}_{nest}$ vector corresponding to the elongation of the superlattice peaks in (a).

Recently, incommensurate superlattice reflections were reported in $Cu_xTiSe_2$ (at ambient pressure) [39] and $1T$-$TiSe_2$ (under pressure) [50]. In our XRD measurement of $Cu_xTiSe_2$ (and pressurized $1T$-$TiSe_2$ to be discussed later), we cannot estimate the degree of incommensurability due to the limited resolution of our IP detector (0.03 deg./pixel). However, both additional superlattice reflections and the change of symmetry suggesting the incommensurate long-range structure were not observed in $Cu_{0.05}TiSe_2$ (and pressurized $1T$-$TiSe_2$).

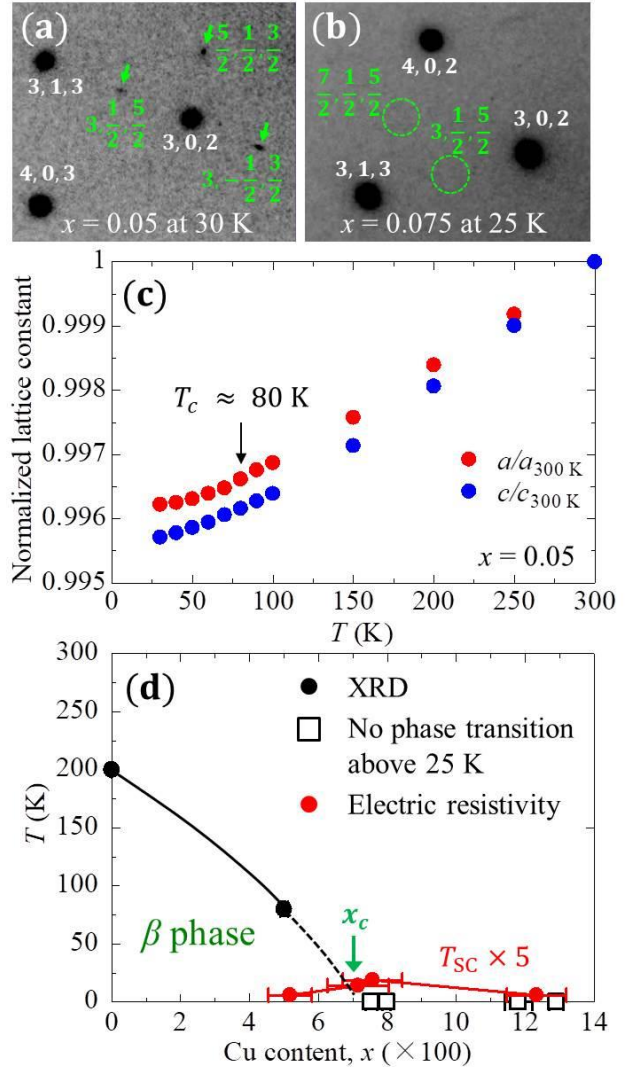

FIG. 6 X-ray diffraction data of (a) $Cu_{0.05}TiSe_2$ at 30 K and (b) $Cu_{0.075}TiSe_2$ at 25 K. (c) Lattice constants normalized to 300 K of $Cu_{0.05}TiSe_2$. (d) $x$–$T$ phase diagram of $Cu_xTiSe_2$. White squares denote that $x$ values for which no superlattice reflections or diffuse scattering are observed above 25 K, indicating absence of any phase transition.

The crystal structure analysis was carried out using the extracted intensity of the superlattice reflections. A triple-$q$ structure identical to pristine $1T$-$TiSe_2$ in the $\beta$ phase was obtained from our analysis (Table I). The size of the dipole moment $|\bm{\mu}|$ decreased to 58% to that observed in $1T$-$TiSe_2$ at 30 K. It is noted that the SC and the triple-$q$ structure coexist in the ground state in $Cu_{0.05}TiSe_2$. In $Cu_{0.05}TiSe_2$, an unusual behavior of lattice constants in the stacking direction due to the structural phase transition is not observed [Fig. 6(c)] as seen in $1T$-$TiSe_2$ (Fig. 2). The compression ratio of $c$ axis was larger than that of $a$ axis, akin to many common layered compounds.

For $0.075 \leq x \leq 0.13$, no additional superlattice reflections or diffuse scattering were observed above 25 K down to $10^{-6}$ times the intensity of the main reflections. For example, Fig. 6(b) shows the XRD data in $Cu_{0.075}TiSe_2$ at 25 K in which there are no superlattice reflections. This fact indicates the absence of structural phase transitions and electric dipole moments.

Figure 6(d) shows the $x$–$T$ phase diagram of $Cu_xTiSe_2$ from our XRD and electric resistivity measurements. The critical concentration $x_c$ of the phase transition is $x \approx 0.07$. It is noted that $x_c$ is inside the SC dome. While this differs from the previous XRD data [39], the present result shows good agreement with those obtained by earlier Raman scattering [42] and IXS [49].

### C. $1T$-$TiSe_2$ at high-pressure

Our $x$–$T$ phase diagram of $Cu_xTiSe_2$ is different from previous reports of the XRD measurement of $Cu_xTiSe_2$ [39] and pressurized $1T$-$TiSe_2$ [50]. Hence, we carefully investigated the crystal structure across the $P$−$T$ phase diagram. To discuss the electronic state under the high-pressure with highly accurate structural parameters, the structure analysis by the synchrotron XRD experiment under the high-pressure were carried out. The intensity of superlattice reflections is approximately three orders of magnitude lower than the intensity of main reflections in the $\beta$ phase of $1T$-$TiSe_2$. Powder samples are often used for structure analysis under pressure, but it is difficult to accurately extract superlattice intensities with weak intensity. Therefore, it is required to perform structure analysis using single crystal samples in this case. However, the accessible reciprocal space is limited due to the constraints imposed by the use of a DAC. It is difficult to perform a comprehensive structure analysis using only one single crystal under high-pressure.

To clarify the crystal structure of $1T$-$TiSe_2$ under pressure, two single crystal samples with different crystal orientations [(i) $30 \times 25 \times 10$ μm$^3$ and (ii) $20 \times 20 \times 10$ μm$^3$] were measured at the same time in the DAC [inset of Fig. 7(a)]. The independent reciprocal space that could be measured was 24% of the total region (resolution limit of $d > 0.75$ Å) with only **crystal 1**, whereas it improved to 69% by using two crystals. The ratio of diffracted intensities from the two crystals was **crystal 1**:**crystal 2** = 1:0.74, in which common reflections were used as the calibration standard.

Figure 7(a) shows the XRD data of $1T$-$TiSe_2$ under high-pressure in the HT phase (3.12 GPa and 110 K). The pink and the black circles correspond to diffraction peaks of **crystal 1** and **crystal 2**, respectively. The information about $a^*b^*$ plane and $c^*$ stacking direction are extracted mainly from **crystal 1** and **crystal 2** respectively. Figure 7(b) shows the measurement points on the $P$–$T$ phase diagram. Crystal structure analyses at the high-pressure HT phase (3.12 GPa and 110 K) and the high-pressure low-temperature phase (2.82 GPa and 5 K) were carried out by using the diffraction intensity from the two crystals. From the analysis of the high-pressure HT phase data, the value of $R_1$ ($I > 4\sigma$) was 4.77%, which is sufficiently reliable (Table I). The volume of the $TiSe_6$ octahedra decreases to 96% (from 22.2081 to 21.4241 Å$^3$) and the distance between the inter-layer Se atoms decreases to 91% [from 2.93678(16) to 2.677(6) Å] at 3.12 GPa and 110 K, when compared to 0 GPa and 215 K. These results are opposite to the structural changes observed with increasing $x$ in $Cu_xTiSe_2$.

Figure 7(c) shows the temperature dependence of the peak profile of a superlattice reflection (−1/2, 7/2, −3/2) from **crystal 1**. The plot point shows the changes in the peak profile as the temperature is lowered from 110 to 5 K at an applied pressure of about 3 GPa [corresponding to the five data points near 3 GPa in Fig. 7(b)]. By decreasing temperature at 3.12 GPa, superlattice reflections appeared below 65 K. This result is consistent with earlier XRD reports [50]. Because all superlattice reflections satisfy the extinction rule of $c$-glide in the $2a \times 2a \times 2c$ lattice, the space group is determined to be $P\bar{3}c1$, the same as the $\beta$ phase. The full-width at half-maximum of the superlattice reflection was almost equal to the main peak (−3, 3, −2) at 5 K [left inset of Fig. 7(c)]. Hence, crystal structure analysis could be performed including the data from the superlattice reflections. At the high-pressure and $\beta$ phase (2.82 GPa and 5 K), the value of $R_1$ ($I > 2\sigma$) was 4.71%, which was comparable to that of the high-pressure HT phase. The similarity of this triple-$q$ structure to the ones in unpressurized $1T$-$TiSe_2$ and $Cu_{0.05}TiSe_2$ was confirmed (Table I). The size of the dipole moment $|\mu|$ was reduced to 56% compared to that of 0 GPa and 30 K. It was confirmed that there were no superlattice reflections at 6.49 GPa and 10 K. This result is consistent with earlier reports [50], which mentioned that superlattice reflections were not observed above $P_c = 5.1$ GPa. The positional relationship between the critical point ($x_c$ and $P_c$) and the SC dome is different between $Cu_xTiSe_2$ and pressurized $1T$-$TiSe_2$. This is also consistent with the previously reported Raman scattering [42,51] and IXS measurements [49].

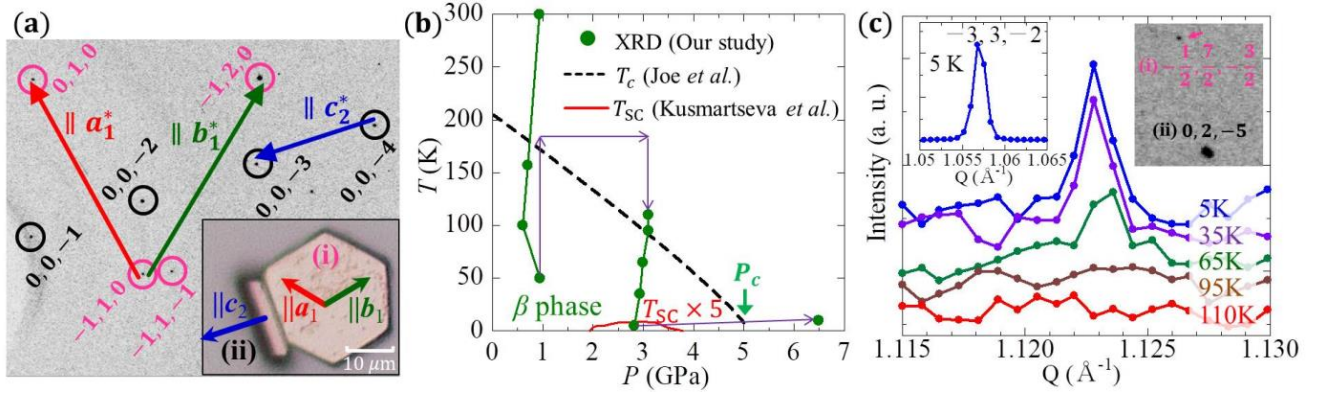

FIG. 7(a) X-ray diffraction data from 1$T$-TiSe$_2$ at 3.12 GPa and 110 K, which is the high-temperature phase. Inset: Two single crystals of 1$T$-TiSe$_2$ with different crystal orientations used in the experiment with a diamond anvil cell. (i) and (ii) indicate **crystal 1** and **crystal 2**, respectively. The pink and black circles show diffraction peaks of **crystal 1** and **crystal 2**, respectively. (b) $P-T$ phase diagram. The green dots show our measurement points. The phase boundary described by the black dotted line and the SC dome described by red line are taken from Ref. [50] and Ref. [38], respectively. (c) The ($-1/2$, $7/2$, $-3/2$) superlattice peak of **crystal 1** for several temperatures, under an applied pressure of around 3 GPa [corresponding to the five points in (b)]. Right Inset: Diffraction peaks of 1$T$-TiSe$_2$ at 2.82 GPa and 5 K. (i) and (ii) indicate ($-1/2$, $7/2$, $-3/2$) of **crystal 1** and (0, 2, $-5$) of **crystal 2**, respectively. Left Inset: The ($-3$, 3, $-2$) peak of the **crystal 1** at 2.82 GPa and 5 K.

TABLE I. Crystallographic data for 1$T$-TiSe$_2$ and Cu$_x$TiSe$_2$.

|  | High-temperature phase | | | | $\beta$ phase | | |
|---|---|---|---|---|---|---|---|
| Pressure (GPa) | 0 | 0 | 0 | 3.12 | 0 | 0 | 2.82 |
| Temperature (K) | 300 | 300 | 300 | 110 | 30 | 30 | 5 |
| Cu content $x$ | 0 | 0.0798(12) | 0.129(2) | 0 | 0 | 0.0500(11) | 0 |
| Wavelength (Å) | 0.35422 | 0.35374 | 0.35374 | 0.4133 | 0.35422 | 0.38814 | 0.4133 |
| Space group | $P\bar{3}m1$ | $P\bar{3}m1$ | $P\bar{3}m1$ | $P\bar{3}m1$ | $P\bar{3}c1$ | $P\bar{3}c1$ | $P\bar{3}c1$ |
| $a (= b)$ [Å] | 3.5395(3) | 3.5447(8) | 3.5481(3) | 3.4763(17) | 7.05930(10) | 7.0597(2) | 6.955(3) |
| $c$ (Å) | 6.0082(4) | 6.0375(9) | 6.0530(4) | 5.7481(17) | 11.9929(3) | 12.0200(8) | 11.524(3) |
| $V$ (Å$^3$) | 65.187(9) | 65.70(2) | 65.992(9) | 60.16(5) | 517.581(17) | 518.81(4) | 482.8(3) |
| $Z$ | 1 | 1 | 1 | 1 | 8 | 8 | 8 |
| $F(000)$ | 90 | 119 | 119 | 90 | 720 | 732 | 720 |
| $d_{min}$ (Å)$^a$ | 0.4 | 0.45 | 0.35 | 0.64 | 0.3 | 0.3 | 0.69 |
| $N_{Tot,obs}$ | 504 | 452 | 5106 | 118 | 108669 | 76833 | 190 |
| $N_{Uniq,obs}$ | 300 | 223 | 663 | 47 | 7344 | 6606 | 76 |
| $R_1$ (%) ($\sigma$ cut) | 2.77 ($I > 4\sigma$) | 1.94 ($I > 4\sigma$) | 2.25 ($I > 4\sigma$) | 4.77 ($I > 4\sigma$) | 2.62 ($I > 2\sigma$) | 2.77 ($I > 2\sigma$) | 4.71 ($I > 2\sigma$) |

| High-temperature phase ($P\bar{3}m1$) | | | | $\beta$ phase ($P\bar{3}c1$) | | | |
|---|---|---|---|---|---|---|---|
| Atom | Site | Position [$x, y, z$] | $B_{eq}$ (Å$^2$) | Atom | Site | Position [$x, y, z$] | $B_{eq}$ (Å$^2$) |
| ($P = 0$ GPa, $T = 300$ K, $x = 0$) | | | | ($P = 0$ GPa, $T = 30$ K, $x = 0$) | | | |
| Se | 2$d$ | [2/3, 1/3, 0.25514(5)] | 0.659(6) | Se(1) | 12$g$ | [0.16401(2), 0.33312(2), 0.12151(4)] | 0.238(2) |
| Ti | 1$a$ | [0, 0, 0] | 0.896(10) | Se(2) | 4$d$ | [2/3, 1/3, 0.12208(5)] | 0.141(3) |
| [$P = 0$ GPa, $T = 300$ K, $x = 0.0798(12)$] | | | | Ti(1) | 6$f$ | [0, 0.49148(3), 1/4] | 0.270(5) |
| Se | 2$d$ | [2/3, 1/3, 0.25399(4)] | 0.592(7) | Ti(2) | 2$a$ | [0, 0, 1/4] | 0.313(8) |
| Ti | 1$a$ | [0, 0, 0] | 0.835(10) | [$P = 0$ GPa, $T = 30$ K, $x = 0.0500(11)$] | | | |
| Cu | 1$b$ | [0, 0, 1/2] | 1.35(11) | Se(1) | 12$g$ | [0.16550(3), 1/3, 0.122098(5)] | 0.1993(2) |
| [$P = 0$ GPa, $T = 300$ K, $x = 0.129(2)$] | | | | Se(2) | 4$d$ | [2/3, 1/3, 0.122098] | 0.1993 |
| Se | 2$d$ | [2/3, 1/3, 0.25339(3)] | 0.708(2) | Ti(1) | 6$f$ | [0, 0.49493(9), 1/4] | 0.3142(2) |
| Ti | 1$a$ | [0, 0, 0] | 0.968(4) | Ti(2) | 2$a$ | [0, 0, 1/4] | 0.3142 |
| Cu | 1$b$ | [0, 0, 1/2] | 1.46(3) | Cu(1) | 6$e$ | [0, 1/2, 0] | 0.47(2) |
| ($P = 3.12$ GPa, $T = 110$ K, $x = 0$) | | | | Cu(2) | 2$b$ | [0, 0, 0] | 0.47 |
| Se | 2$d$ | [2/3, 1/3, 0.2671(6)] | 0.24(7) | ($P = 2.82$ GPa, $T = 5$ K, $x = 0$) | | | |
| Ti | 1$a$ | [0, 0, 0] | 0.31(9) | Se(1) | 12$g$ | [0.16486(16), 1/3, 0.1169(2)] | 0.16(7) |
|  |  |  |  | Se(2) | 4$d$ | [2/3, 1/3, 0.1169] | 0.16 |
|  |  |  |  | Ti(1) | 6$f$ | [0, 0.4953(4), 1/4] | 0.24(9) |
|  |  |  |  | Ti(2) | 2$a$ | [0, 0, 1/4] | 0.24 |

$^a d_{min}$ indicates the resolution limit used for the crystal structure analysis. $N_{Tot,obs}$ and $N_{Uniq,obs}$ indicate the number of the total reflections and the unique reflections in the $d_{min}$ region, respectively. In $P = 0$ GPa, $T = 30$ K, $x = 0.0500(11)$, to reduce the parameters, we restricted $y[Se(1)] = 1/3$ and $z[Se(1)] = z[Se(2)]$ and $B_{eq}[Se(1)] = B_{eq}[Se(2)]$ and $B_{eq}[Ti(1)] = B_{eq}[Ti(2)]$ and $B_{eq}[Cu(1)] = B_{eq}[Cu(2)]$. In $P = 2.82$ GPa, $T = 5$ K, $x = 0$, to reduce the parameters, we restricted $y[Se(1)] = 1/3$ and $z[Se(1)] = z[Se(2)]$ and $B_{eq}[Se(1)] = B_{eq}[Se(2)]$ and $B_{eq}[Ti(1)] = B_{eq}[Ti(2)]$.

## IV. DISCUSSION

The phase transition of 1$T$-TiSe$_2$ is suppressed by pressure and electron doping. These behaviors resemble to those of a general CDW transition in low-dimensional materials. The suppression of the CDW transition can be explained by changes in band filling and dimensionality of the system. Changes in EPC and excitonic interaction due to external pressure and carrier doping in 1$T$-TiSe$_2$ system have been discussed in several experiments [12,20,22-25] and calculations [17-19,21]. From our structural findings, the nature of the phase transition in this system becomes clearer as follows.

As the inter-layer Se-Se distance shrinks [$h_{inter}(P = 3.12)/h_{inter}(P = 0) \approx 0.91$, $dh_{inter}/dP \approx -0.083$ Å/GPa] due to increasing pressure in the HT phase, the inter-layer interaction in 1$T$-TiSe$_2$ becomes strong. The volume of TiSe$_6$ octahedra decreases [$V(P = 3.12)/V(P = 0) \approx 0.96$, $dV/dP \approx -0.25$ Å$^3$/GPa] on applying pressure. In the $P-T$ phase diagram of other TMDs such as 1$T$-TaS$_2$ [3], 2$H$-NbSe$_2$ [5], and 1$T$-TiTe$_2$ [68], the SC is stabilized for a wide range of pressures. On the other hand, the critical pressure $P_c$ occurs in the higher-pressure region rather than the SC dome in pressurized 1$T$-TiSe$_2$ [Fig. 7(b)]. From our results under pressure, $2a \times 2a \times 2c$ structure is realized within the SC dome. The triple-$q$ structures were observed even around the SC dome in pressurized 1$T$-TiSe$_2$, while the size of the electric dipole moments $|\boldsymbol{\mu}|$ was reduced to 56% compared to the $\beta$ phase at ambient pressure. This result suggests the possibility that the excitonic interaction and SC coexist.

Recently, the similar $P-T$ phase diagram and SC was reported in Ta$_2$NiSe$_5$ [69,70], which is another excitonic insulator candidate. Electric toroidal moments are also observed in the excitonic phase of Ta$_2$NiSe$_5$ [26]. However, Ta$_2$NiSe$_5$ is a ferrotoroidic state, while 1$T$-TiSe$_2$ is an antiferrotoroidic state [Fig. 1(e)]. These similarities between 1$T$-TiSe$_2$ and Ta$_2$NiSe$_5$ are important factors in the fundamental understanding of these compounds with excitonic interactions. The excitonic interaction and SC may be closely related in both 1$T$-TiSe$_2$ and Ta$_2$NiSe$_5$ under applied pressure.

The balance between the number of electrons and holes is an important consideration in the context of excitonic interactions. Because the band dispersion itself does not essentially change in Cu$_x$TiSe$_2$ [48,71], a positive shift of the chemical potential occurs by electron doping. In this case, because hole carriers are reduced by electron doping, the excitonic interaction is weakened. From our structural studies of Cu$_x$TiSe$_2$, as the inter-layer Se-Se distance increases [$h_{inter}(x = 0.08)/h_{inter}(x = 0) \approx 1.010$, $dh_{inter}/dx \approx 0.35$ Å/$x$] with increasing $x$ in the HT phase, the inter-layer interaction in 1$T$-TiSe$_2$ becomes weak. This affects the temperature dependence of the lattice constants [Figs. 2 and 6(c)], in which the anomalous behavior of the compression ratio of $c$ axis disappears by Cu intercalation. The volume of TiSe$_6$ octahedra increases [$V(x = 0.08)/V(x = 0) \approx 1.003$, $dV/dx \approx 0.91$ Å$^3$/$x$] by Cu intercalation. These are opposite to the changes caused by pressure.

With regard to the $x-T$ phase diagram, $x_c$ exists within the SC dome [Fig. 6(d)], which is different from the position of $P_c$ in the $P-T$ phase diagram. This difference may be related to the presence or absence of the excitonic interaction in 1$T$-TiSe$_2$. On the other hand, the structural change accompanying the $\beta$ phase transition in TiSe$_2$ layers suppressed by Cu intercalation is similar to that by pressure. These atomic displacements in the $\beta$ phase cannot be explained by changes in the spatial charge disproportionation because there is no difference in the volume of the two types of TiSe$_6$ octahedra as mentioned above. For this reason, the $\beta$ phase of 1$T$-TiSe$_2$ cannot be described by a standard CDW framework.

Recently, studies from our group have confirmed the pure CDW due to electron-electron nesting on the high Cu-doped region ($x \approx 0.33$) [71], in which there are two kinds of TiSe$_6$ octahedra with different volumes from each other. In the CDW state in Cu$_{0.33}$TiSe$_2$, the charge disproportionation occurs in TiSe$_2$ layers and there is no excitonic interaction because of no hole pockets. This pure CDW state is different from that in the $\beta$ phase. This result also seems to indicate that not only the EPC but also the excitonic interaction are important in the $\beta$ phase of 1$T$-TiSe$_2$.

## V. SUMMARY

We investigated the crystal structure of pristine 1$T$-TiSe$_2$ under ambient- and high-pressure and Cu$_x$TiSe$_2$ under ambient-pressure by using synchrotron XRD. In pristine 1$T$-TiSe$_2$ in the triple-$q$ CDW state, the characteristic antiferroelectric arrangement of the electric dipoles and the possible electric toroidal moments was discussed. The structural changes are significantly different between Cu intercalation and pressure application in the HT phase. This result implies that the pressure and carrier doping effects to physical parameters of 1$T$-TiSe$_2$ are different from each other. Furthermore, because the crystal structures around the SC state are different between Cu$_x$TiSe$_2$ and pressurized 1$T$-TiSe$_2$, SC in the two scenarios may have different origins. Our structural study gives valuable information about the phase transition and the SC in this system, the larger theoretical implications of which remain to be understood.

## ACKNOWLEDGMENTS

We thank T. Kaneko, S. Murakami, S. Ishihara, and T. Hasegawa for fruitful discussions. This work was supported by a Grant-in-Aid for Scientific Research (Grants No. JP23244074, and No. JP17K17793) from JSPS, the Kato Foundation for Promotion of Science, and the Daiko Foundation. The synchrotron radiation experiments were performed at SPring-8 with the approval of the Japan Synchrotron Radiation Research Institute (JASRI) (Proposal No. 2013B0083, No. 2016B1270, No. 2016B3783, and No. 2017B1733). A part of this work was performed under the Shared Use Program of QST Facilities (Proposal No.2016B-H10) supported by QST Advanced Characterization Nanotechnology Platform under the remit of "Nanotechnology Platform" of the Ministry of Education, Culture, Sports, Science and Technology (MEXT), Japan (Proposal No. A-16-QS-0023). In addition, a part of this work was carried out under the Visiting Researcher's Program of the Institute for Solid State Physics, the University of Tokyo, and the Institute for Molecular Science (IMS), supported by Nanotechnology Platform Program (Molecule and Material Synthesis) of the MEXT, Japan.

[†]Present address: Department of Physics, Nagoya University, Nagoya 464-8602, Japan.
[‡]Present address: Japan Synchrotron Radiation Research Institute, SPring-8, 1-1-1 Kouto, Sayo, 679-5198, Japan.
[§]Present address: Synchrotron Radiation Research Center, National Institutes for Quantum and Radiological Science and Technology, Sayo, Hyogo 679-5148, Japan.
[#]Present address: Department of Chemistry, Rikkyo University, Toshima-ku, Tokyo 171-8501, Japan.

[1] J. G. Bednorz, and K. A. Müller, Z. Phys. B **64**, 189 (1986).
[2] C. B. Scruby, P. M. Williams, and G. S. Parry, Phil. Mag. **31**, 255 (1975).
[3] B.Sipos, A. F. Kusmartseva, A. Akrap, H. Berger, L. Ferró, and E. Tutiš, Nat. Mater. **7**, 960 (2008).
[4] D. E. Moncton, J. D. Axe, and F. J. DiSalvo, Phys. Rev. Lett. **34**, 734 (1975).
[5] C. Berthier, P. Molinié, and D. Jérome, Solid State Commun. **18**, 1393 (1976).
[6] T. Rohwer, S. Hellman, M. Wiesenmayer, C. Sohrt, A. Stange, B. Slomski, A. Carr, Y. Liu, L. M. Avila, M. Kalläne, S. Mathias, L. Kipp, K. Rossnagel, and M. Bauer, Nature (London) **471**, 490 (2011).
[7] E. Möhr-Vorobeva, S. L. Johnson, P. Beaud, U. Staub, R. De Souza, C. Milne, G. Ingold, J. Demsar, H. Schaefer, and A. Titov, Phys. Rev. Lett. **107**, 036403 (2011).
[8] M. Porer, U. Leierseder, J. M. Ménard, H. Dachraoui, L. Mouchliadis, I. E. Perakis, U. Heinzmann, J. Demsar, K. Rossnagel, and R. Huber, Nat. Mater. **13**, 857 (2014).
[9] L. J. Li, E. C. T. O'Farrell, K. P. Loh, G. Eda, B. Özyilmaz, and A. H. Castro Neto, Nature (London) **529**, 185 (2016).
[10] Q. Yao, D. W. Shen, C. H. P. Wen, C. Q. Hua, L. Q. Zhang, N. Z. Wang, X. H. Niu, Q. Y. Chen, P. Dudin, Y. H. Lu, Y. Zheng, X. H. Chen, X. G. Wan, and D. L. Feng, Phys. Rev. Lett. **120**, 106401 (2018).
[11] J. Ishioka, Y. H. Liu, K. Shimatake, T. Kurosawa, K. Ichimura, Y. Toda, M. Oda, and S. Tanda, Phys. Rev. Lett. **105**, 176401 (2010).
[12] A. Kogar, M. S. Rak, S. Vig, A. A. Husain, F. Flicker, Y. I. Joe, L. Venema, G. J. MacDougall, T. C. Chiang, E. Fradkin, J. van Wezel, P. Abbamonte, Science **358**, 1314 (2017).
[13] A. Zunger, and A. J. Freeman, Phys. Rev. B **17**, 1839 (1978).
[14] F. J. Di Salvo, D. E. Moncton, and J. V. Waszczak, Phys. Rev. B **14**, 4321 (1976).
[15] N. Suzuki, A. Yamamoto, and K. Motizuki, J. Phys. Soc. Jpn. **54**, 4668 (1985).
[16] R. Bianco, M. Calandra, and F. Mauri, Phys. Rev. B **92**, 094107 (2015).
[17] T. Kaneko, Y. Ohta, and S. Yunoki, Phys. Rev. B **97**, 155131 (2018).
[18] Y. Yoshida, and K. Motizuki, J. Phys. Soc. Jpn. **49**, 898 (1980).
[19] M. Holt, P. Zschack, H. Hong, M.Y. Chou, and T.-C. Chiang, Phys. Rev. Lett. **86**, 3799 (2001).
[20] F. Weber, S. Rosenkranz, J.-P. Castellan, R. Osborn, G. Karapetrov, R. Hott, R. Heid, K.-P. Bohnen, and A. Alatas, Phys. Rev. Lett. **107**, 266401 (2011).
[21] M. Hellgren, J. Baima, R. Bianco, M. Calandra, F. Mauri, and L. Wirtz, Phys. Rev. Lett. **119**, 176401 (2017).
[22] D. Jérome, T. M. Rice, and W. Khon, Phys. Rev. **158**, 462 (1967).
[23] Th. Pillo, J. Hayoz, H. Berger, F. Lévy, L. Schlapbach, and P. Aebi, Phys. Rev. B **61**, 16213 (2000).
[24] T. E. Kidd, T. Miller, M. Y. Chou, and T.-C. Chiang, Phys. Rev. Lett. **88**, 226402 (2002).
[25] H. Cercellier, C. Monney, F. Clerc, C. Battaglia, L. Despont, M. G. Garnier, H. Beck, P. Aebi, L. Patthey, H. Berger, and L. Forró, Phys. Rev. Lett. **99**, 146403 (2007).

[26] A. Nakano, T. Hasegawa, S. Tamura, N. Katayama, S. Tsutsui, and H. Sawa, Phys. Rev. B **98**, 045139 (2018).
[27] K. Okazaki, Y. Ogawa, T. Suzuki, T. Yamamoto, T. Someya, S. Michimae, M. Watanabe, Y. Lu, M. Nohora, H. Takagi, N. Katayama, H. Sawa, M. Fujisawa, T. Kanai, N. Ishii, J. Itatani, T. Mizokawa, and S. Shin, Nat. Commun. **9**, 4332 (2018).
[28] Y. Wakisaka, T. Sudayama, K. Takubo, T. Mizokawa, M. Arita, H. Namatame, M. Taniguchi, N. Katayama, M. Nohara, and H. Takagi, Phys. Rev. Lett. **103**, 026402 (2009).
[29] T. Kaneko, T. Toriyama, T. Konishi, and Y. Ohta, Phys Rev. B **87**, 035121 (2013).
[30] Y. Murakami, D. Golež, M. Eckstein, and P. Werner, Phys. Rev. Lett. **119**, 247601 (2017).
[31] Y. F. Lu, H. Kono, T. I. Larkin, A. W. Rost, T. Takayama, A. V. Boris, B. Keimer, and H. Takagi, Nat. Commun. **8**, 14408 (2017).
[32] S. Mor, M. Herzog, D. Golež, P. Werner, M. Eckstein, N. Katayama, M. Nohara, H. Takagi, T. Mizokawa, C. Monney, and J. Stähler, Phys. Rev. Lett. **119**, 086401 (2017).
[33] D. Werdehausen, T. Takayama, M. Höppner, G. Albrecht, A. W. Rost, Y. Lu, D. Manske, H. Takagi, and S. Kaiser, Sci. Adv. **4**, eaap8652 (2018).
[34] E. Morosan, H. W. Zandbergen, B. S. Dennis, J. W. G. Bos, Y. Onose, T. Klimczuk, A. P. Ramirez, N. P. Ong, and R. J. Cava, Nat. Phys. **2**, 544 (2006).
[35] E. Morosan, K. E. Wagner, L. L. Zhao, Y. Hor, A. J. Williams, J. Tao, Y. Zhu, and R. J. Cava, Phys. Rev. B **81**, 094524 (2010).
[36] N. Giang, Q. Xu, Y. S. Hor, A. J. Williams, S. E. Dutton, H. W. Zandbergen, and R. J. Cava, Phys. Rev. B **82**, 024503 (2010).
[37] K. Sato, T. Noji, T. Hatakeda, T. Kawamata, M. Kato, and Y. Koike, J. Phys. Soc. Jpn. **86**, 104701 (2017).
[38] A. F. Kusmartseva, B. Sipos, H. Berger, L. Forró, and E. Tutiš, Phys. Rev. Lett. **103**, 236401 (2009).
[39] A. Kogar, G. A. de la Pena, S. Lee, Y. Fang, S. X.-L. Sun, D. B. Lioi, G. Karapetrov, K. D. Finkelstein, J. P. C. Ruff, P. Abbamonte, and S. Rosenkranz, Phys. Rev. Lett. **118**, 027002 (2017).
[40] A. M. Novello, M. Spera, A. Scarfato, A. Ubaldini, E. Giannini, D. R. Bowler, and Ch. Renner, Phys. Rev. Lett. **118**, 017002 (2017).
[41] S. Yan, D. Iaia, E. Morosan, E. Fradkin, P. Abbamonte, and V. Madhavan, Phys. Rev. Lett. **118**, 106405 (2017).
[42] H. Barath, M. Kim, J. F. Karpus, S. L. Cooper, P. Abbamonte, E. Fradkin, E. Morosan, and R. J. Cava, Phys. Rev. Lett. **100**, 106402 (2008).
[43] G. Wu, H. X. Yang, L. Zhao, X. G. Luo, T. Wu, G. Y. Wang, and X. H. Chen, Phys. Rev. B **76**, 024513 (2007).
[44] G. Li, W. Z. Hu, J. Dong, D. Qian, D. Hsieh, M. Z. Hasan, E. Morosan, R. J. Cava, and N. L. Wang, Phys. Rev. Lett. **99**, 167002 (2007).
[45] D. Qian, D. Hsieh, L. Wray, E. Morosan, N. L. Wang, Y. Xia, R. J. Cava, and M. Z. Hasan, Phys. Rev. Lett. **98**, 117007 (2007).
[46] J. F. Zhao, H.W. Ou, G. Wu, B. P. Xie, Y. Zhang, D.W. Shen, J. Wei, L. X. Yang, J. K. Dong, M. Arita, H. Namatame, M. Taniguchi, X. H. Chen, and D. L. Feng, Phys. Rev. Lett. **99**, 146401 (2007).
[47] T. Jeong, and T. Jarlborg, Phys. Rev. B **76**, 153103 (2007).
[48] R. A. Jishi and H. M. Alyahyaei, Phys. Rev. B **78**, 144516 (2008).
[49] M. Maschek, S. Rosenkranz, R. Hott, R. Heid, M. Merz, D. A. Zocco, A. H. Said, A. Alatas, G. Karapetrov, S. Zhu, J. van Wezel, and F. Weber, Phys. Rev. B **94**, 214507 (2016).
[50] Y. I. Joe, X. M. Chen, P. Ghaemi, K. D. Finkelstein, G. A. de la Peña, Y. Gan, J. C. T. Lee, S. Yuan, J. Geck, G. J. MacDougall, T. C. Chiang, S. L. Cooper, E. Fradkin, and P. Abbamonte, Nat. Phys. **10**, 421 (2014).
[51] C. S. Snow, J. F. Karpus, S. L. Cooper, T. E. Kidd, and T.-C. Chiang, Phys. Rev. Lett. **91**, 136402 (2003).
[52] M. Calandra, and F. Mauri, Phys. Rev. Lett. **106**, 196406 (2011).
[53] A. A. Titov, A. I. Merentsov, A. E. Kar'kin, A. N. Tiov, and V. V. Fedorenko, Phys. Solid State **51**, 230 (2009).
[54] K. Sugimoto, H. Ohsumi, S. Aoyagi, E. Nishibori, C. Moriyoshi, Y. Kuroiwa, H. Sawa, and M. Takata, AIP Conf. Proc. **1234**, 887 (2010).
[55] T. Watanuki, A. Machida, T. Ikeda, A. Ohmura, H. Kaneko, K. Aoki, T. J. Sato, and A. P. Tsai, Philos. Mag. **87**, 2905 (2007).
[56] C.-S. Zha, H.-K. Mao, and R. Hemley, Proc. Natl. Acad. Sci. U.S.A. **97**, 13494 (2000).
[57] K. Sugawara, K. Sugimoto, T. Fujii, T. Higuchi, N. Katayama, Y. Okamoto, and H. Sawa, J. Phys. Soc. Jpn. **87**, 024601 (2018).
[58] R. H. Blessing, Crystallogr. Rev. **1**, 3 (1987).
[59] V. Petříček, M. Dušek, and L. Palatinus, Z. Kristallogr. Cryst. Mater. **229**, 345 (2014).
[60] A. B. Holder, and H. Büttner, J. Phys. Condens. Matter **14**, 7973 (2002).
[61] R. H. Buttner, and E. N. Maslen, Acta Crystallogr., Sect.B: Struct. Sci., Cryst. Eng. Mater. **48**, 764 (1992).
[62] Yu V Kopaev, Phys.-Usp. **52**, 1111 (2009).
[63] L. Y. Guo, M. H. Li, Q. W. Ye, B. X. Xiao, and H. L. Yang, Eur. Phys. J. B **85**, 208 (2012).
[64] A. Planes, T. Castán, and A. Saxena, Philos. Mag. **94**, 1893 (2014).
[65] S. Hayami, H. Kusunose, and Y. Motome, Phys. Rev. B **90**, 024432 (2014).

[66] G. A. Wiegers, Physica B **99**, 151 (1980).
[67] S. H. Huang, G. J. Shu, W. W. Pai, H. L. Liu, and F. C. Chou, Phys. Rev. B **95**, 045310 (2017).
[68] U. Dutta, P. S. Malavi, S. Sahoo, B. Joseph, and S. Karmakar, Phys. Rev. B **97**, 060503(R) (2018).
[69] A. Nakano, K. Sugawara, S. Tamura, N. Katayama, K. Matubayashi, T. Okada, Y. Uwatoko, K. Munakata, A. Nakao, H. Sagayama, R. Kumai, K. Sugimoto, N. Maejima, A. Machida, T. Watanuki, and H. Sawa, IUCrJ **5**, 158 (2018).
[70] K. Matubayashi, N. Katayama, R. Yamanaka, A. Hisada, T. Okada, A. Nakano, H. Sawa, K. Munakata, A. Nakao, T. Kaneko, T. Toriyama, T. Konishi, Y. Otha, H. Okamura, T. Mizokawa, M. Nohara, H. Takagi, and U. Uwatoko (unpublished).
[71] S. Kitou, S. Kobayashi, T. Kaneko, N. Katayama, S. Yunoki, T. Nakamura, and H. Sawa, Phys. Rev. B **99**, 081111(R) (2019).